# Re-visiting the Distance Coefficient in Gravity Model
## Haonan Wu

**Abstract**

**This paper revisits the classic gravity model in international trade and reexamines the distance coefficient. As pointed out by Frankel (1997), this coefficient measures the relative unit transportation cost between short distance and long distance rather than the absolute level of average transporation cost. Our results confirm this point in the sense that the coefficient has been very stable between 1991-2006, despite the obvious technological progress taken place during this period. Moreover, by comparing the sensitivity of these coefficients to change in oil prices at short periods of time, in which technology remained unchanged, we conclude that the average technology has indeed reduced the average trading cost. The results are robust when we divide the aggregate international trades into different industries.**

## Introduction

In social science, gravity models are used to predict and describe certain behaviors that mimic gravitational interaction as described in Isaac Newton's law of gravity. Generally, the social science models contain some elements of mass and distance, which lends them to the metaphor of physical gravity. In recent economic literature, the Gravity model has been given new meaning. It's about trade in international economics, similar to other gravity models in social science, predicts bilateral `trade flows` based on the economic sizes (often using GDP measurements) and distance between two units. The model was first used by Tinbergen in 1962. The basic model for trade between two countries ( i and j) takes the form of:

$$F_{ij} = G(M_i^{\beta_1} M_j^{\beta_2} / D_{ij}^{\beta_3})$$

where F is the amount of trade between country i and country j, measured by its value and G is a function of their masses and distance. The model has also been used in international relations to evaluate the impact of treaties and alliances on trade, and it has been used to test the effectiveness of trade agreements and organizations such as the North American Free Trade Agreement (NAFTA) and the World Trade Organization (WTO). Jeffery Frankel and Ernesto Stein have tested this impact in *Trade blocs and the Americas: The natural, the unnatural and the super-natural (1993).* They drew the conclusion from their study that some degree of preferences along natural continental lines, such as the Free Trade Area of the Americas or enlargement of the European Union to include EFFA and Eastern Europe, would be a good thing, but that the formation of Free Trade Areas where the preferences approach 100% would represent an excessive degree of regionalization of world trade.   Grossman and Helpman (1995) even did some further researches to analyze the reasons why free-trade zone can be founded. They examined in *The Politics of Free-Trade Agreements* that the conditions under which a free-trade agreement might emerge as an equilibrium outcome of a negotiation between politically-minded governments. Both the political benefit and the political cost are measured by a weighted sum of the change in industry profits and the change in average welfare in going from the status quo to bilateral free trade. The weights on benefits in one country and costs in the other reflect the negotiating abilities of the two governments (i.e., the "Nash weights") and the political welfare that would accrue to the two governments if they rejected the agreement entirely.

The model has been an empirical success in that it accurately predicts trade flows between countries for many goods and services, but for a long time some scholars believed that there was no theoretical justification for the gravity equation. However, a gravity relationship can arise in almost any trade model that includes trade costs that increase with distance. For example, Deardorff (2004) once provided important amendments to Ricardian model which contains trade costs. He proved that the net trade of one industry (no matter bilateral or global) also depends on the cost of production and trade costs of one country relative to other countries. Here transportation cost certainly plays a necessary role. Judging from his econometrical results, when the distance between two countries is farther, trade cost which is mainly consisted of transportation cost is always higher. Then following this phenomenon, the bilateral trade becomes less, which is consistent with the gravity relationship.

While the model's basic form consists of factors that have more to do with geography and spatiality, the gravity model has been used to test hypotheses rooted in purer economic theories of trade as well. One such theory predicts that trade will be based on relative factor abundances. One of the common relative factor abundance models is the Heckscher-Ohlin model. This theory would predict that trade patterns would be based on relative factor abundance. Those countries with a relative abundance of one factor would be expected to produce goods that require a relatively large amount of that factor in their production. While a generally accepted theory of trade, many economists in the Chicago School believed that the Heckscher-Ohlin model alone was sufficient to describe all trade, while Bertil Ohlin himself argued that in fact the world is more complicated. Investigations into real world trading patterns have produced a number of results that do not match the expectations of comparative advantage theories. Notably, a study by Wassily Leontief found that the United States, the most capital endowed country in the world, actually exports more in labor-intensive industries. Comparative advantage in factor endowments would suggest the opposite would occur. Other theories of trade and explanations for this relationship were proposed in order to explain the discrepancy between Leontief's empirical findings and economic theory. The problem has become known as the Leontief paradox.

Past research using the gravity model has also sought to evaluate the impact of various variables in addition to the basic gravity equation. Among these, price level and exchange rate variables have been shown to have a relationship in the gravity model that accounts for a significant amount of the variance not explained by the basic gravity equation. According to empirical results on price level, the effect of price level varies according to the relationship being examined. For instance, if exports are being examined, a relatively high price level on the part of the importer would be expected to increase trade with that country. A non-linear system of equations are used by Anderson and van Wincoop (2003) to account for the endogenous change in these price terms from trade liberalization. A more simple method is to use a first order log-linearization of this system of equations (Baier and Bergstrand (2009)), or exporter-country-year and importer-country-year dummy variables. For counterfactual analysis, however, one would still need to account for the change in world prices.

**Estimation of Gravity Equations**

Since the gravity model for trade does not hold exactly, in econometric applications it is customary to specify

$$F_{ij} = G(M_i^{\beta_1} M_j^{\beta_2} / D_{ij}^{\beta_3}) \eta_{ij},$$

where $F_{ij}$ represents volume of trade from country $i$ to country $j$, $M_i$ and $M_j$ typically represent the GDPs for countries $i$ and $j$, $D_{ij}$ denotes the distance between the two countries, and $\eta$ represents an error term with expectation equal to 1.

The traditional approach to estimating this equation consists in taking logs of both sides, leading to a log-log model of the form (note: constant G becomes part of $\beta_0$):

$$\ln(F_{ij}) = \beta_0 + \beta_1 \ln(M_i) + \beta_2 \ln(M_j) - \beta_3 \ln(D_{ij}) + \epsilon_{ij}.$$

However, this approach has two major problems. First, it obviously cannot be used when there are observations for which $F_{ij}$ is equal to zero. Second, it has been argued by Santos Silva and Tenreyro (2006) that estimating the log-linearized equation by least squares (OLS) can lead to significant biases. As an alternative, these authors have suggested that the model should be estimated in its multiplicative form, i.e,

$$F_{ij} = \exp[\beta_0 + \beta_1 \ln(M_i) + \beta_2 \ln(M_j) - \beta_3 \ln(D_{ij})] \eta_{ij},$$

using a Poisson pseudo-maximum likelihood (PPML) estimator usually used for count data (see the original paper for details). One of the authors' more surprising findings was that, when controlling for sharing a common language, having past colonial ties does not increase trade. This is despite the fact that simpler methods, such as taking simple averages of trade shares of countries with and without former colonial ties suggest that countries with former colonial ties continue to trade more. Santos Silva and Tenreyro (2006) did not explain where their result came from and even failed to realize their results were highly anomalous. Martin and Pham (2008) argued that using PPML on gravity severely biases estimates when zero trade flows are frequent. However, their results were challenged by Santos Silva and Tenreyro (2011), who argued that the simulation results of Martin and Pham (2008) are based on misspecified models and showed that the PPML estimator performs well even when the proportions of zeros is very large.

In applied work, the model is often extended by including variables to account for language relationships, tariffs, contiguity, access to sea, colonial history, exchange rate regimes, and other variables of interest.

The two most important factors of the gravity model in explaining bilateral trade flows are the geographical distance between the two countries, and their economic size. Indeed, these two variables give the gravity model its name. A large part of the apparent bias toward intra-regional trade is certainly due to simple geographical proximity. Indeed Krugman (1991b) suggests that most of it may be due to proximity, so that the three trading blocs are welfare-improving 'natural' groupings. Despite the obvious importance of distance and transportation costs in determining the volume of trade, empirical studies surprisingly often neglect to measure this factor. The measure taken by Frankel (1993) is the log of distance between the two major cities (usually the capital) of the respective countries. In detail, the different measures of distance involve border distance, capital city distance and major city distance weighted by relative economic size. Generally, people regard capital as the representative of one country. It is the most politically important city to a country as all of trade decisions are made there. Thus in the academic research of international trade, scholars are accustomed to choosing capital city distance as the main measure of distance variable. Some other professors like Helpman advocated selecting major cities as the factor. A major city seems closer to the country's economic center of gravity (Chicago for the United States rather than Washington DC, and Shanghai for China rather than Beijing). The entire economic activity of a large country is concentrated at a single point of mass. Given this point, major city distance weighted by relative economic size seems to be a better choice for our research. They also add a dummy 'Adjacent' variable to indicate when two countries share a common land border.

Entering GNPs in product form is empirically well-established in bilateral trade regressions. It can be justified by the modern theory of trade under imperfect competition. In addition there is reason to believe that GNP per capita has a positive effect on trade, for a given size: as countries become more developed, they tend to specialize more and to trade more. The equation to be estimated, in its most basic form, is

$$\log(T_{ij}) = \alpha + \beta_1 \log(GNP_i GNP_j) + \beta_2 \log(GNP/pop_i GNP/pop_j) \\ + \beta_3 \log(DISTANCE_{ij}) + \beta_4(ADJACENT_{ij}) + \gamma_1(EA_{ij}) \\ + \gamma_2(EC_{ij}) + \gamma_3(NAFTA_{ij}) + u_{ij}. \qquad (1)$$

*EA, EC,* and *NAFTA* are three of the dummy variables we use when testing the effects of membership in a common regional grouping standing for East Asia, European Community, and North America.

To most readers who have no enough knowledge on trade theory, the assumption that trade between countries depends positively on their size and inversely on distance may seem self-evident. Although the derivation of a proportionate relationship between trade flows and country size is an important foundation, the theories of Helpman (1987) and other authors cited do not include a role for distance and thus cannot be called theories of the full gravity model. There are also a few imperfect-substitutes theory, which incorporated a role for only transportation costs, proxied in practice by distance.

A widespread perception hold the view that the current wave of globalization, much like the first, should have led to the "death of distance." Other things equal, globalization should generate a dispersion of economic activity reflecting a decline in transaction costs, especially transport costs. But studies based on the traditional gravity model of international trade—the workhorse for studies on the pattern of trade and the influence of transport costs—do not reach that conclusion. Most former researches relied on distance as a proxy for transport costs, obtaining an estimated elasticity of bilateral trade with respect to distance in the range [_1.3; _0.8]. However, as will be detailed, when the model is estimated separately for several years, the absolute value of the coefficient almost always increases over time. This is puzzling, because the common perception of globalization is that distance should be becoming less important in international trade, implying decreasing rather than increasing values for the estimated coefficient of distance. Jean-Franc¸ois Brun, Ce´line Carre`re, Patrick Guillaumont, and Jaime de Melo (2002) argued that several variants of gravity model were used to address the distance puzzle for a sample of 130 countries over the period 1962–96. The puzzle proved robust to several ad hoc versions of the gravity model, but it was significantly reduced when the gravity model was correctly specified to include remoteness (or an index of multilateral trade resistance). Adding an augmented trade barrier function (real price of oil, index of infrastructure, and share of primary exports in total bilateral trade) that corrects for the misspecification inherent in the standard representation of transport costs by distance yielded plausible estimates of the expected death of distance.

Despite the many shortcomings associated with gravity-based indirect estimates of transport costs, several intuitively plausible results emerge from the model estimations: an elasticity of trade to income close to unity, a significant impact of the real exchange rate on the volume of bilateral trade, and expected significant signs for exporter and importer country characteristics and for the impact of remoteness on the volume of trade.

In recent years, a few further researches about the extension of the distance in gravity model have appeared. In Regional trading Blocs, the author admitted that transport costs will not always and everywhere be monotonically increasing in distance, let alone in a convenient logarithmic form. The author argued that when the adjacency variable is not included in the equation, the estimated coefficient on the log of distance is about -.75 (Frankel 1993). This means that when the distance between two countries is increased by 1.0 percent, trade between them falls by about three-quarters of a percent. However the adjacency variable should be included. The Netherlands is close to France and Korea to Japan, but without the common border the effect is not the same. The controlling for adjacency tends to get lower coefficients on the log of distance.

Bikker (1987) measured distance by sea routes, which tries a clever way of isolating the role of physical shipping costs from the other costs of doing business at a distance. Although he added a variable for the additional sea distance that had to be covered between the country pair, divided by the normal distance, he finally concluded that physical shipping costs are less important than conventionally assumed, considering the low estimated coefficient. Here he did not do any further researches which involved other transportation cost such as air or railway.

In addition, the former research has tried disaggregating trade into three categories. The results show higher distance effects for manufacturers than for agricultural products or other raw materials. All findings confirm that physical transport costs are not necessarily the most important component of costs associated with distance.

Claudia M. Buch, Jörn Kleinert, and Farid Toubal (2003) drew the conclusion in their research that increasing volumes of global trade and capital flows are indicators of the globalization of the world economy. Deregulation and technological progress are likely to have lowered the costs of bridging large distances and to have led to a decline in 'distance costs'. Beyond this conventional wisdom, economists are interested in empirically assessing the magnitude of these changes. Since direct measures of distance costs are often unavailable, geographic distance between countries is often used as a proxy. Many applications of gravity equations suggest that the coefficient on distance has not changed significantly over time, and this could be taken as evidence against declining distance costs. They have argued that this interpretation of distance coefficients is misleading. Essentially, people cannot infer changes in distance costs from changes in distance coefficients obtained from cross-section equations for different years. In the extreme case of a proportional decline in distance costs and a proportional increase in bilateral economic linkages, the effects of changes in distance costs would show up solely in the constant term of gravity equations.

These considerations do not imply, of course, that distance coefficients are uninformative with regard to globalization trends. Falling distance costs do have caused a strong increase in international activities of all kinds. Hence, the often pro-claimed 'death of distance' has not occurred, and distance is still an important determinant of international economic activity. However, the correct interpretation of constant distance coefficients is that international activities between countries that are located far away from each other and between countries that are located close to each other have expanded at similar proportions.

**The result for empirical analysis**

In the empirical analysis, I have tried disaggregating the trade into three categories: Agriculture trade, Manufacturing trade and Service trade. The basic equation to be estimated is:

$log(T_{ij})\{log(Agri\text{-}Trade_{ij}), log(Manu\text{-}Trade_{ij}), log(Serv\text{-}Trade_{ij})\} = j + \beta_1 log(GDP_i) + \beta_2 log(GDP_j) + \beta_3 log(DISTANCE_{ij}) + \beta_4 log(ADJACENT_{ij}) + \beta_5 log(LANGUAGE) + \beta_6 log(FREE\text{-}TRADE) + \beta_7 log(POPULATION_i) + \beta_8 log(POPULATION_j) + \delta$

The left side of the equation is total trade between country i and j. There are five variables on the right side: Distance between two countries, GDP of country i and GDP of country j. The other three are dummy variables: Same Language ij, Free Trade ij and Adjacent. These two variables are equal to 1 when countries i and j share the same language and belong to the same free trade zone, such as NAFTA or FTA. Otherwise, they are equal to 0. The empirical analysis focuses on the distance coefficient. Agriculture trade, Manufacturing trade and Service trade are used to replace the total trade respectively in order to see whether the regression result of distance coefficient varies for different products. Consistent with our theories, the results not only differ in absolute values, but also in the correlation with oil prices. The time series we choose is from 1991 to 2006. Here we choose 10 countries that are mainly from OECD: United States, China, Germany, Russia, South Africa, Norway, Brazil, Italy, Austria and Canada. These countries are distributes in five continents which were separated by oceans and land. They are paired with each other to form 45 bilateral trade ties. The geographical distance between two countries is measured by the distance between capitals. The data of GDP and trade amount are mainly from WTO and OECD online-Library network. The data of capital distance is collected from mileage database of China's major airline corporations. All of our statistics including the time period we choose is well-fitted to our analysis. First, the time period we choose is from 1991 to 2006, almost 16 years in which technological progress had taken place in this relatively long period. Secondly, given the fact that oil played a necessary role in transporation cost from 1991 to 2006, the average annual oil price is a valid proxy for transportation cost per unit distance.

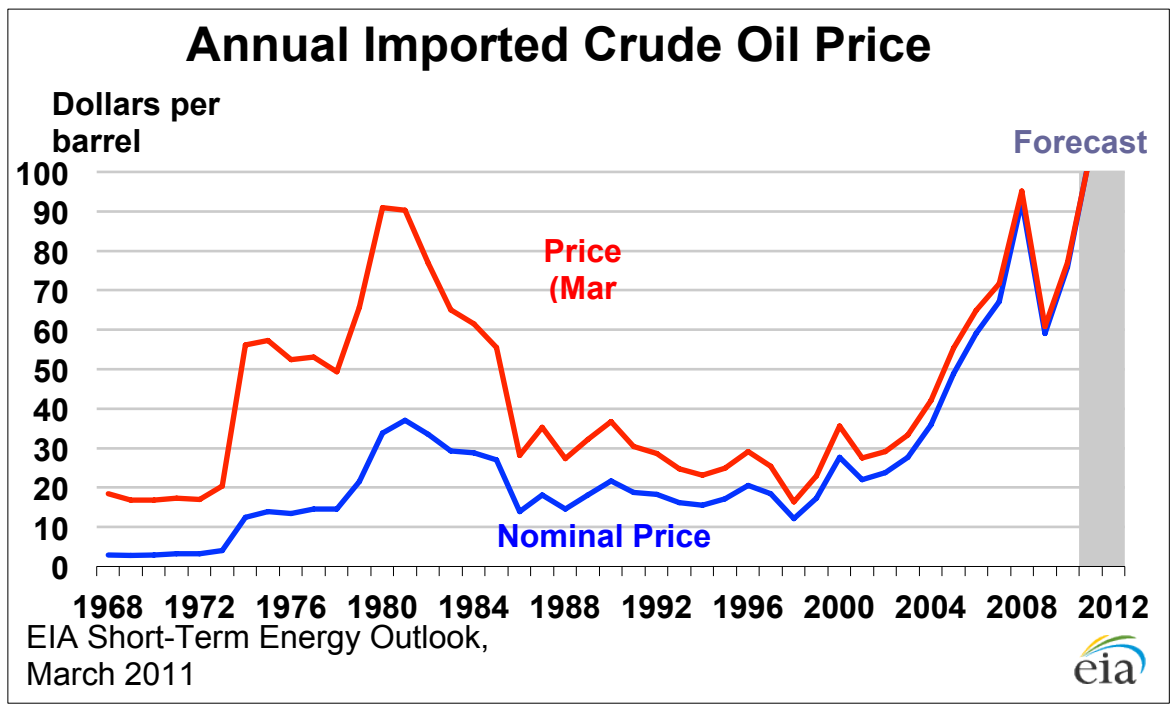

**Agricultural distance coefficient:**

Judging from Chart 1, we can see that from 1991 to 2006, the agricultural distance coefficient has switched sign from being positive to being negative. The overall trend is downward. Especially from 1991 to 1998, the trend is more significant but still above x-axis, which means geographical distance and agricultural trade between two countries showing positive relationship during this period.  The larger the geographical distance, the more agricultural trade

took place between these two countries. One example of Agricultural Trade regression result is as follows:

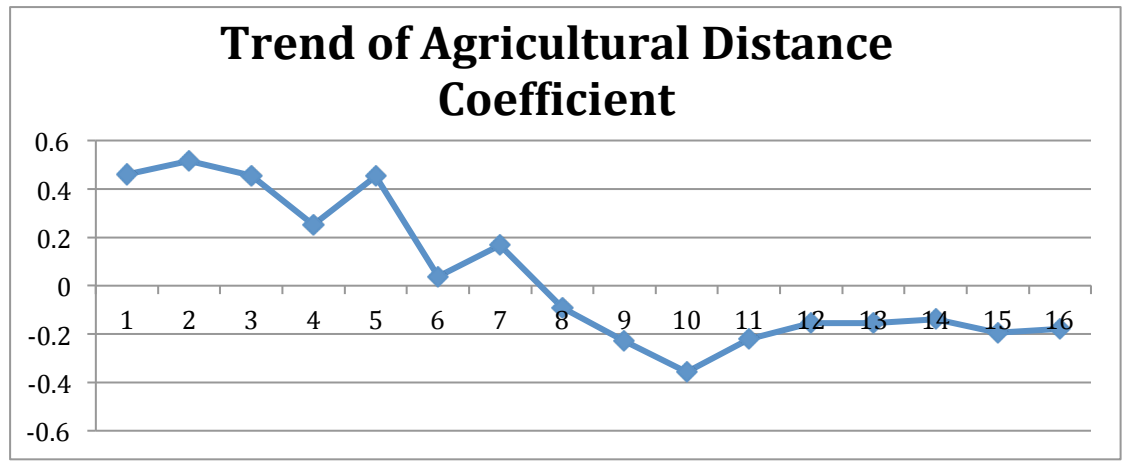

---

-> year = 1995

| log_Agri | Coef. | Std. Err. | t | P>|t| | [95% Conf. Interval] |
|---|---|---|---|---|---|
| log_gdpa | 1.189409 | .3434068 | 3.46 | 0.001 | .4907426   1.888 |
| log_gdpb | .7843716 | .2613699 | 3.00 | 0.005 | .2526105   1.316 |
| log_dis | .4539043 | .3793289 | 1.20 | 0.240 | -.31788   1.225 |
| adj | .2747158 | 1.74993 | 0.16 | 0.876 | -3.285544   3.835 |
| samelanguage | .3666536 | .9043762 | 0.41 | 0.688 | -1.473313   2.206 |
| free-tradezone | 2.639705 | 1.507364 | 1.75 | 0.089 | -.4270504   5.706 |
| populationA | -.0000552 | .0000459 | -1.20 | 0.237 | -.0001485   .00003 |
| populationB | 1.06e-07 | 5.34e-06 | 0.02 | 0.984 | -.0000108   .00001 |
| _cons | -19.10029 | 6.913944 | -2.76 | 0.009 | -33.16682   -5.034 |

---

This result deviates from the general theory of gravity model, which states that trade between two countries and their geographical distance should be inversely correlated. Taking the type of products traded in this particular industry into account in the 1990s. Science and technology were not well developed and transportation methods are limited to land and sea transportation. Distance coefficient, which represents the transportation cost in the real trade is not as important because the marginal cost per distance for sea and land transportation is low. In terms of this background, in order to explain the strange phenomenon that is mentioned above, we need to consider other factors determine the amount of agricultural trade among countries. The difference in climate, which governs the types of agricultural products, is probably the most important one. When the distance between the two trading partners is close, these two countries probably have very similar or even the same climates. Therefore farmers in these two counties will grow similar plants what thrive well under the particular climate. There will not be any comparative advantage over the other country in growing the species. There will very little, if any agriculcural trade changes hand across border. [i]

Conversely, the greater geographical distance between two countries is, the more likely they belong to different types of climates. Differences in the structure of agricultural output will also occur. As a result, agricultural trade will be greater. Based on the analysis above and combined with the Chart 1, we can conclude that during the early stage when technology is not well developed, climate is the dominating factor that determines relation between distance and agricultural trade.

Since 1998, the coefficient has turned into negative. By 2006 it has been negative, but remain smoothly fluctuating around zero. In the 21st century, the rapid development of science and technology made important contributions to the agricultural trade between the countries. Transportation methods become diversified and the distance coefficient is more sensitive. The explanation for this is as follows:

In the late 1900s, due to the limited transportation method, the transport of agricultural products between countries often choose those which have a longer shelf life. These products were shipped by large container ships. The marginal transportation cost is low and hence the coefficient on distance is dominated by climate diversity. However, the development of technology brings diversify to transport methods. Countries are able to do trade of wider range of agricultural products among each other. Some fresh products with shorter shelf life can be transported through more efficient methods. These methods include high speed train or airplanes, which have high transportation cost per unit distance. The distance coefficient is becoming increasingly sensitive. Transportation costs start to become a major factor affecting agricultural distance coefficient instead of climate. Considering general conclusions of the gravity model, transportation costs has negative effects on distance coefficient. Still we are unable to completely ignore the positive effect of climate on coefficient. Therefore, because of the simultaneous positive and negative effects, the net value of the coefficient remains smoothly fluctuating around zero.

**Industrial distance coefficient:**

Industry is quite different from agriculture. Technology in agricultural products has very small differences between countries, and therefore it is difficult to generate comparative advantages under similar geographic conditions. But when it comes to manufacturing, each country will have comparative advantage in their own specialized industries. According to Hecksher－Ohlin model, different locations of countries determine their various initial endowments of industrial raw materials, which will then lead to the formation of comparative advantages. In Ricardian model, it is the difference in technology that brings the difference in comparative advantage. Because the physical properties of industrial products are relatively stable and the production is in general not affected by natural climatic and other external factors, in the discussion of industrial distance

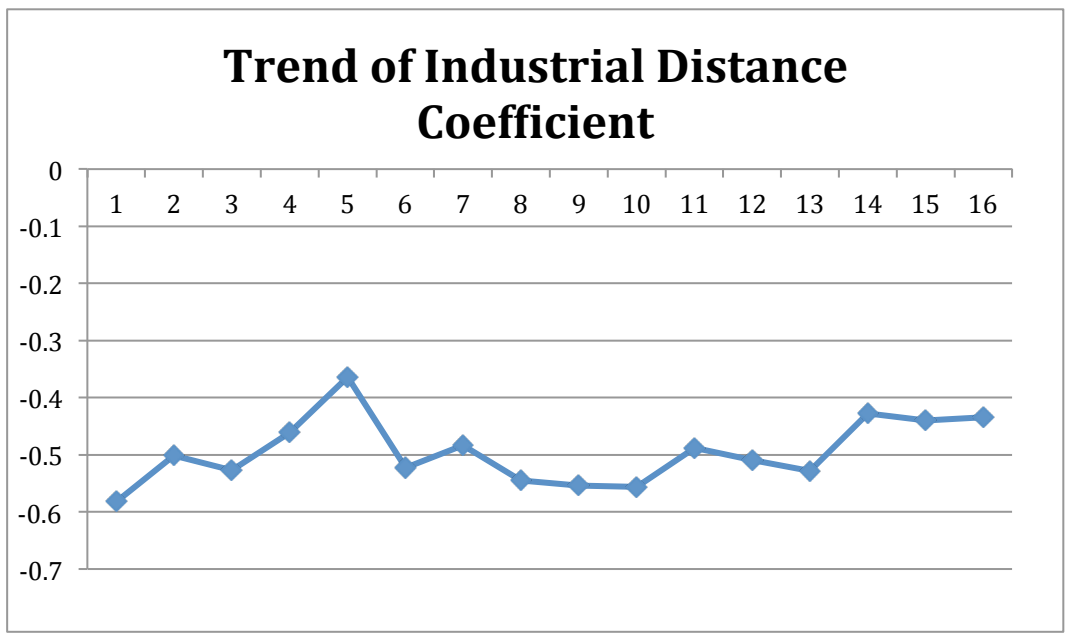

coefficient, the physical distance becomes important variable in quantitative analysis. An example of Industrial Trade regression result is:

----------------------------------------------------------------------------------------------------------

-> year = 1995

| log_man | Coef. | Std. Err. | t | P>|t| | [95% Conf. Interval] |

```
log_gdpa       1.364882    .1336502    10.21    0.000     1.092969    1.637
log_gdpb       .8500027    .1017223     8.36    0.000      .6430471   1.057
log_dis |     -.3644761    .1476307    -2.47    0.019     -.6648329   -.0641
adj     |      .5868481    .6810538     0.86    0.395     -.7987663   1.972
samelanguage   .156862     .3519733     0.45    0.659     -.5592332    .873
freetradezone  .8352486    .5866497     1.42    0.164     -.3582992   2.029
populationa   -.0000583    .0000178    -3.27    0.003     -.000094    -.00002
populationb    6.90e-06    2.08e-06     3.32    0.002      2.67e-06    .00001
_cons         -11.72632    2.690832    -4.36    0.0001    -17.20086   -6.252
```
-------------------------------------------------------------------------------

In general, judging from Chart 2, from 1991 to 2006, the industrial distance coefficient remains a stabilizing fluctuation around -0.5 as a whole. But in some specific years, changes are still significant. Given this, we divided our discussion into two parts: the global pattern and the local pattern. First, we observe that there are significant changes in certain years that are correlated with the change in oil price. Judging from Chart2 and Chart3, from 1993 to 1995, oil prices rose slightly, up from $ 23.71 per barrel to $ 24.89 per barrel. From 1995 to 1998, oil prices were falling from $ 24.89 per barrel to $ 16.38 per barrel. Corresponding to two special periods, industrial distance coefficient is also undergoing similar changes.

From 1993 to 1995, the oil prices increased corresponding with the distance

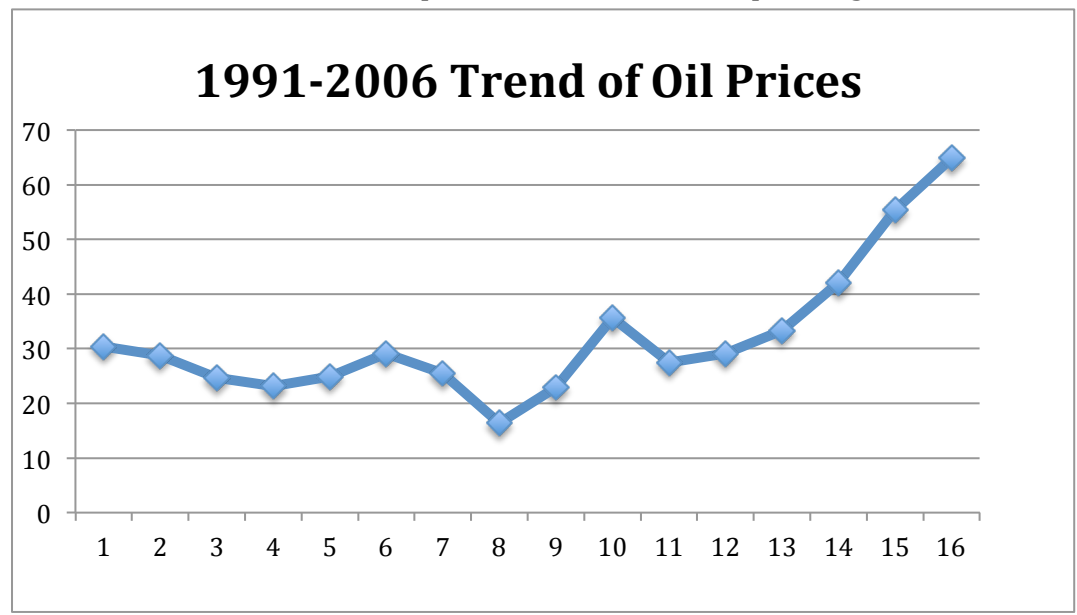

1991-2006 Trend of Oil Prices

coefficient, but this relationship is not significantly. From 1995 to 1998, during this period, crude oil prices fell quickly accompanied by a significant decline in industrial distance coefficient. Such a special phenomenon indicates the distance coefficient may be relatively sensitive to changes in oil prices.

On the other hand, considering the entire time series, as mentioned above, the industrial distance coefficient remains a stabilizing fluctuation around -0.5. Especially since 1998, Chart3 shows a continued rise in crude oil prices, from $ 16.38 climbing to $ 64.83 per barrel. Even if crude oil prices have increased so significantly, changes in the distance coefficient is not large and the overall trend is relatively stable. In the analysis of agricultural distance coefficient we mentioned earlier, the development of science and technology has brought a diversity of transport methods. Moreover it also has a great impact on the changes of the industrial distance coefficient. Since 1998, although oil price is rising, because of the development of science and technology, big progress in the efficiency of using fuel, the appearance of a series of renewable resources and so on affect the sensitivity of the distance coefficient to crude oil prices. To summarize, in the role of science and technology evolving, despite a significant increase in crude oil prices, changes in industrial distance coefficient is not large. The more developed of technology, the lower the sensitivity of distance coefficient is to oil price.

Frankel (1993) mentioned his point in ***The Gravity Model of the Bilateral Trade*** that in the long run, the distance coefficient do not has a great change with the development of science and technology. He believes that the average changing trend of distance coefficient is unable to effectively reflect the development and progress of science and technology. From empirical analysis of industrial distance coefficient, it can be seen that this conclusion is reasonable. Frankel thinks that technological progress reduced transportation costs at all distances by some fixed percentage of their previous level. Then there would be no reason for the coefficient on log distance to fall. In other words, no matter how far or how close between two countries, their decline rates of transportation costs are similar. So in this case, the distance coefficient will not have upward trend, along with the advancement of technology. In the long run it will always exhibit a relatively stable situation.

**Service distance coefficient:**

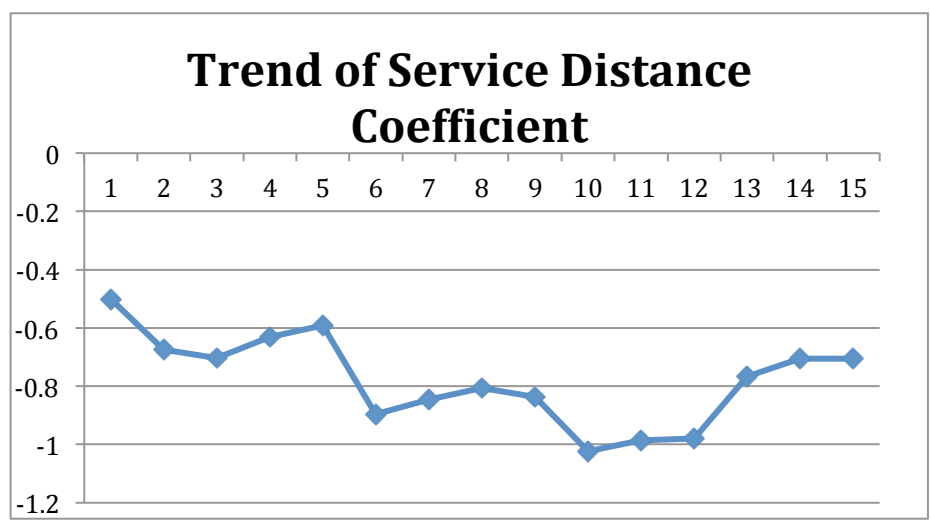

From 1991 to 2006, crude oil prices as a whole have a rising trend. Service distance coefficient in contrast has a slowly decline trend. Unlike agricultural trades, which requires different transportation methods for different types of agricultural products, trade in service are more homogeneous and hence exhibit similar patterns to that in manufacturing. The average transportation cost for trade in service is quite sensitive to the changes of crude oil prices. One example of Service trade regression result is as follows:

-------------------------------------------------------------------------------------------------------

→ year = 1995

| log_service | Coef. | Std. Err. | t | P>|t| | [95% Conf. Interval] | |
|---|---|---|---|---|---|---|
| log_gdpa | 1.362347 | .3300309 | 4.13 | 0.000 | .6908942 | 2.03 |
| log_gdpb | .5766517 | .2511894 | 2.30 | 0.028 | .0656029 | 1.088 |
| log_dis | -.5910588 | .3645539 | -1.62 | 0.114 | -1.332749 | .1506 |
| adj | .4287094 | 1.68177 | 0.25 | 0.800 | -2.992877 | 3.850 |
| samelanguage | .3280344 | .8691503 | 0.38 | 0.708 | -1.440265 | 2.096 |
| freetradezone | .4338249 | 1.448652 | 0.30 | 0.766 | -2.513479 | 3.381 |
| populationa | -.000079 | .0000441 | -1.79 | 0.082 | -.000168 | .00001 |
| populationb | -1.38e-07 | 5.14e-06 | -0.03 | 0.979 | -.000011 | .00001 |
| cons | -8.124555 | 6.644643 | -1.22 | 0.230 | -21.64318 | 5.394 |

-------------------------------------------------------------------------------------------------------

Now let us go back to Frankel's point of view: the average trend of distance coefficient does not reflect the impact of technological advancement. An important premise is that Frankel believes in the long term, no matter how far or how close between two countries, their decline rates of transportation costs are similar. We found that in this conclusion, Frankel mainly focused on the perspective of technology and did not take into changes in the price of crude oil which has impact on transport costs occured. From Chart2, we find that from 1993 to 1998 along with the decline and the rise in oil prices, the industrial distance coefficient has a great fluctuation. It is very sensitive to oil price changes. Since nearly a decade later in 1998, despite a substantial increase in crude oil prices, the coefficient always maintains in the range of -0.4 to -0.5. It becomes insensitive. So we can conclude that the problem left behind in the study of Frankel can be well explained by *the sensitivity analysis between distance coefficient to crude oil prices.* Although the changes in long term of distance coefficient does not reflect progress in technology, the sensitivity between distance coefficient to the changes in crude oil prices could prove impact of development of science and technology on gravity model. The more advanced technology is, the lower the sensitivity of distance coefficient is to oil prices. The relatively stable developing trend after 1998 from empirical analysis well confirms this conclusion.

**Conclusion**

The conclusion of this study involves three main points as follows. From the beginning, we confirm that judging from the results of regression in our time series: 1990—2006, with the development of science and technology, there is no obvious tendency for the effect of relative distance to fall. Thus we believe Frankel's point is correct: the average trend of the change of distance coefficient is unable to reflect the advancement of technology. He gives us a possible explanation that if the technological progress reduced shipping costs at all distances by some fixed percentages of their previous level, then there would be no reason for the coefficient on log distance to fall. Then in this essay, we try to give a reasonable explanation of the impact of technological progress on international trade. Considering the regression of our model and regarding the industrial distance coefficient as an example, in the role of science and technology evolving, despite a significant increase in crude oil prices from 1998 to 2006, changes in industrial distance coefficient is not large. The more developed of technology, the lower the sensitivity of distance coefficient is to oil price. This kind of sensitivity can be applied to Agricultural distance coefficient and Service distance coefficient as well. That is to say this sensitivity can reflect the technological advancement effectively. The last point is that we tried to disaggregate total distance coefficient into three categories: Agriculture, Manufacture and Service in order to do our further researches. When we did regressions separately, three different trends of distance coefficient appeared. Sharp changes occur in agricultural distance coefficient during the period from 1991 to 2006. While sudden shock appeared in our analysis of Manufacturing and Service distance coefficient. We tried to give reasonable explanations to various trends of each industry. Judging from our researches of each distance coefficient above, obviously we get some interesting phenomenon. We believe this further analysis of distance coefficient will certainly contribute to the study of gravity model and it will give people a better understanding of international trade.

[i] According to the theory in the Ricardian model, comparative advantage is the cause of bilateral trade. According to the Washington Council on International Trade, comparative advantage is the ability to produce a good at a lower cost, relative to other goods, compared to another country. In the *Principles of Economics*, Ricardo states that comparative advantage is a specialization technique used to create more efficient production and describes opportunity cost between producers. With perfect competition and undistorted markets, countries tend to export goods in which they have a comparative advantage. For example, we should think of two countries that both make cards and pencils and use the same amount of time to make one unit of items. Country one can make 4 pencils if they specialize just in pencils at the expense of one card, but this country can also make ¼ of a card at the expense of one pencil. The same logic goes for country two: if country two makes only pencils, it will make 2 pencils at the expense of 1 card. If country two specializes only in cards, it will make ½ of a card at the expense of a pencil. For this example, country one has a comparative advantage in pencils over country two (4 pencils to 2 pencils), whereas, country two has a comparative advantage in cards over country one (½ of a card to ¼ of a card). In Ricardo's idea of comparative advantage, these two countries should specialize in what they do best.